\documentclass[prd,aps,floats,floatfix,superscriptaddress,preprintnumbers,
showpacs,eqsecnum,nofootinbib,twocolumn]{revtex4}
\usepackage{latexsym,array,theorem,mathrsfs,bm,float}

\usepackage{psfrag}
\usepackage{amsfonts,amsmath,amssymb,latexsym,array,afterpage,
theorem,mathrsfs,bm,float,epsfig,color,graphicx,tabularx,here,multirow}
\usepackage[dvipdfmx]{hyperref}
\newcommand{\bea}{\begin{eqnarray}}
\newcommand{\ena}{\end{eqnarray}}
\newcommand{\beann}{\begin{eqnarray*}}
\newcommand{\enann}{\end{eqnarray*}}

\newcommand{\ma}[1]{\mbox{$\mathcal{#1}$}}

\newcommand{\ti}{\tilde}

\newcommand{\calhR}[1]{\raisebox{2ex}{\tiny ({\em h})}\hspace{-0.8em}{\ma R}}
\newcommand{\pd}{\partial}
\newcommand{\BS}{\boldsymbol}
\newcommand{\MC}{\mathcal}

\newcommand{\MB}{\mathbb}

\newif\iffigure
\figurefalse

\usepackage{remreset}	



\begin{document}

\title{
Role of Mann Counterterm in Gravitational Energy
}


\author{Shoichiro {\sc Miyashita}}
\email{s-miyashita''at''aoni.waseda.jp}
\address{Department of Physics, Waseda University, 
Okubo 3-4-1, Shinjuku, Tokyo 169-8555, Japan}


\date{\today}

\begin{abstract}
\begin{center}
\bf{Abstract}
\end{center} 
In 1999, R. B. Mann proposed a counterterm that is some sort of generalization of the well-known Holographic counterterm and that can eliminate the divergence of the gravitational action of asymptotically AdS and flat spacetimes (Phys. Rev. D $\BS{60}$ (1999) 104047 \cite{Mann}). I show it is not only for eliminating the divergence of such spacetimes but also for setting the ground state energy to zero for any $d$-dimensional spacetimes with an $S^{d-2} \times \MB{R} $ boundary geometry, and speculate it is also true for spacetimes with any (suitable) boundary geometry and topology.
\end{abstract}


\maketitle

\section*{Proper Gravitational Action}
For a field theory defined on a $(d-1)$ dimensional manifold $\MC{B}$ with a fixed background pseudo-Riemannian geometry $\BS{\gamma}$, there is a ambiguity of adding a term
\bea
\int_{\MC{B}} d^{d-1} y \sqrt{-\gamma} f(\BS{\gamma}) ~ , \label{Counterterm}
\ena
where $f$ is a local function of geometric scalar quantites such as $\MC{R}, \MC{R}_{ij}\MC{R}^{ij}$. For example, consider a scalar field theory
\bea
I_{scalar}[\varphi,\BS{\gamma}] \hspace{6.0cm} \notag \\
= \int_{\MC{B}} d^{d-1} y \sqrt{-\gamma} \left[ -\frac{1}{2} \gamma^{ij}\pd_{i}\varphi \pd_{j}\varphi -V(\varphi) \right] \hspace{0.3cm}
\ena
with a nonzero potential minimum $\displaystyle V_{min}= \min_{\Phi \in \MB{R}} V(\Phi) \neq 0$.
If we add the term
\bea
I_{scalar,ct}[\BS{\gamma}]=\int_{\MC{B}} d^{d-1} y \sqrt{-\gamma} V_{min} ~ ,
\ena 
the ground state energy computed from the total action
\bea
I_{scalar, proper}[\varphi,\BS{\gamma}]=I_{scalar}[\varphi,\BS{\gamma}]+I_{scalar,ct}[\BS{\gamma}]
\ena
is now set to zero for any (suitable) $\MC{B}$ and $\BS{\gamma}$
\footnote{
This simple subtraction works only classically. Quantum mechanically, a slight modification of $I_{scalar,ct}$ is needed, that depends on more finer information of $V$, in addition to $V_{min}$, and that of $\BS{\gamma}$.
}
.
 I call such a action {\it the proper action}.

The objective of this latter is to raise and pursue the possibility that adding Mann counterterm $I_{Mann}[\BS{\gamma}]$ \cite{Mann} 
\footnote{
A closely related work was done by Lau at almost the same time \cite{Lau}.
}
(see eq. (\ref{MannTerm1}), (\ref{MannTerm2}), (\ref{MannTerm3})) to Einstein-Hilbert-York-Gibbons-Hawking action \cite{York, GibbonsHawking} leads to {\it the proper gravitational action} 
\bea
I_{GR, proper}[\BS{g}] = I_{EH}[\BS{g}] + I_{YGH}[\BS{g}]+I_{Mann}[\BS{\gamma}] \hspace{0.5cm} \label{theProperAction} \\
I_{EH}[\BS{g}]= \frac{1}{16\pi G} \int_{\MC{M}} d^{d}x \sqrt{-g} (\MC{R}^{(d)}-2\Lambda) \notag \\
I_{YGH}[\BS{g}]= \frac{1}{8\pi G} \int_{\MC{B}} d^{d-1} y \sqrt{-\gamma} \Theta \hspace{1.2cm} \notag  
\ena
where $\MC{B}$ is time-like boundary(ies) of $\MC{M}$, $\BS{\gamma}$ is a boundary metric that admits a time-like Killing vector. I neglect other boundaries and joints in (\ref{theProperAction}). If this is the case, the ground state energy given by Brown-York tensor \cite{BrownYork} of the action
\bea
\tau_{ij}^{proper} = \frac{-2}{\sqrt{-\gamma}} \frac{\delta I_{GR, proper}[\BS{g}]}{\delta \gamma^{ij}}
\ena
is zero for any suitable boundary geometry $(\MC{B}, \BS{\gamma})$, and for any value of $\Lambda$, at least, at semi-classical level.

The first thing I have to mention about Mann counterterm is that it is equivalent to the well-known Holographic counterterm $I_{ct}[\BS{\gamma}]$ \cite{BalasubramanianKraus, EmparanJohnsonMyers},  completely for even spacetime dimension (as stated in \cite{Mann}), and up to a finite contribution for odd dimension when $\Lambda<0$ and $\BS{\gamma}$ is of infinite volume. 
This means that Mann counterterm can subtract all IR divergences for asymptotically AdS spacetimes (or, in terms of AdS/CFT \cite{Maldacena, GKP, Witten}, all UV divergences of the holographic CFT \cite{HenningsonSkenderis, SusskindWitten} ). Apparently, this is the property that a counterterm of the gravitational action should have.
\footnote{
For subtracting the divergence of asymptotically AdS spacetime, a different sort of counterterm, so called {\it Kounterterm } \cite{Olea1, Olea2}, is known, which works for the boundary condition fixing $K^{i}_{j}$ (in my notation, $\Theta^{i}_{j}$). 
Throughout this letter, I only consider Dirichlet type boundary condition fixing the boundary metric (and ``conformal'' Dirichlet boundary condition fixing the conformal equivalence class of the boundary metric and setting $\tau^{proper}=0$). The relation between this work and \cite{Olea1, Olea2, AnastasiouMiskovicOleaPapadimitriou} or the implication for other types of boundary condition such as \cite{CompereMarolf, KrishnanRaju, KrishnanMaheshwariBalaSubramanian} is left for future work. } 
The next is, contrary to $I_{ct}[\BS{\gamma}]$, Mann counterterm can also be used to eliminate the divergence of asymptotically flat spacetimes of 4- and 5-dimension \cite{KrausLarsenSiebelink}. 

What is mentioned about Mann counterterm newly in this letter is, as stated above, that it is not only for eliminating divergences of the gravitational action or gravitational energy of spacetimes with a boundary of infinite spatial volume, but also for setting the ground state energy to zero, at least for spacetimes with some restricted class of $(\MC{B}, \BS{\gamma})$, and hopefully, for those with any suitable $(\MC{B}, \BS{\gamma})$, namely, the action (\ref{theProperAction}) is proper. Explicitly, after introducing Mann counterterm in a little detail, I show the ground state energy of spacetimes with an $S^{d-2}\times \MB{R}$ boundary geometry of any radius and any value of $\Lambda$ is zero (for $4\leq d \leq 7$). This result indicates that, since these spacetimes are main targets in gravitational thermodynamics \cite{GibbonsHawking, HawkingPage, York2, BrownComerMartinezMelmedWhitingYork}, Mann counterterm is useful for such a situation and we do not have to worry about the choice of counterterm any more
\footnote{
Although it is useful for such a simple situation, the background subtraction method itself is somewhat conceptually unsatisfactory for setting the ground state energy to zero (also for just subtracting the divergence) because we need to know the ground state solution $\BS{g}^{gs}$ (or a meaningful reference state solution $\BS{g}^{ref}$) {\it a priori}. The advantage of the counterterms that depend only on $\BS{\gamma}$, such as $I_{ct}[\BS{\gamma}]$, is the counterterms themselves have the information of the ground states energy (or that of a meaningful reference state) and we do not need to know $\BS{g}^{gs}$ ($\BS{g}^{ref}$). Unfortunately, $I_{ct}[\BS{\gamma}] $ is not defined for $\Lambda\geq 0$ and it does not have the correct information of the ground state energy of spacetimes with a finite radius $S^{d-2} \times \MB{R}$ boundary even for $\Lambda<0$. The ground state energy computed by using $I_{ct}[\BS{\gamma}]$ deviates from zero and its deviation depends on the radius of $S^{d-2}$.
}
.
I end the letter with some remarks on Mann counterterm and the proper gravitational action.

\section*{Mann Counterterm}
Mann counterterm\footnotemark[5]
is constructed out purely from geometric quantities that refer to the information of the boundary metric $\BS{\gamma}$, as is the Holographic counter term $I_{ct}[\BS{\gamma}]$, 
\bea
I_{Mann}[\BS{\gamma}]= \frac{-1}{8\pi G}\int_{\MC{B}}d^{d-1}y \sqrt{-\gamma} \Theta_{Mann}(\BS{\gamma}) \hspace{1.7cm} \label{MannTerm1}
\ena
\bea
\Theta_{Mann}(\BS{\gamma})= \hspace{5.9cm} \notag \\ \begin{cases}
    \displaystyle \hspace{2.9cm} \frac{1}{l} \hspace{2.3cm}  d=3 ~\\
 \\
    \displaystyle \hspace{1.3cm} \left[C_{1}(d)\MC{R} - C_{2}(d) \Lambda\right]^{\frac{1}{2}} \hspace{1.cm}  d=4,5 \\
    ~~ ~~ \\
    \displaystyle \left[D_{1}(d)\left(\MC{R}_{ij} \MC{R}^{ij}+\frac{d-7}{2(d-2)}\MC{R}^2\right) \right.  ~ \\
   \hspace{1.5cm}  - D_{2}(d) \Lambda \MC{R} + D_{3}(d)\Lambda^2  \biggl]^{\frac{1}{4}}   ~  d=6,7  \label{MannTerm2}
\end{cases}
\ena
where the positive coefficients $C_{a}(d), D_{a}(d)$ are
\bea
C_{1}(d)= \frac{d-2}{d-3} ~ , \hspace{1.4cm} C_{2}(d)= \frac{2(d-2)}{d-1}~ , \hspace{0.9cm} \notag \\
D_{1}(d)= \frac{2(d-2)^3}{(d-5)(d-3)^2}~ , D_{2}(d)= \frac{4(d-2)^2}{(d-3)(d-1)},  \label{MannTerm3} \\
 D_{3}(d)= \frac{4 (d-2)^2}{(d-1)^2} ~ . \hspace{4.65cm} \notag 
\ena
Note that $l$ denotes the AdS radius and the $d=3$ case is special, works only for $\Lambda<0$, and is completely same as the Holographic counterterm \cite{BalasubramanianKraus}. Therefore, I will not consider the $d=3$ case here. 
We can easily check that\\
$\cdot$ it coincides with $I_{ct}[\BS{\gamma}]$, completely for even dimension, and up to finite contribution for odd dimension, when we take large $l$ or small curvature limit
\footnotetext[5]{
Precisely, the original counterterm proposed in \cite{Mann} is only for 4-dimension case. The cases of $5\leq d \leq 7$ that I show below are the natural extension of it (the $d=5$ case was also shown in \cite{KrausLarsenSiebelink}).
}
\footnotemark[6]\footnotetext[6]{
The reason why I show only up to 7-dimension case is just $I_{ct}[\BS{\gamma}]$ is explicitly shown only up to 7-dimension in \cite{EmparanJohnsonMyers}. But the derivation of higher dimensional case is straightforward if we know the Holographic counterterm of the dimension as I will comment later.
}
\footnotemark[7]\footnotetext[7]{
For small $x$, $(1+ax)^{\frac{1}{2}} \simeq 1+\frac{1}{2}ax- \frac{1}{8}a^2 x^2$ and $(1+ax+bx^2)^{\frac{1}{4}} \simeq 1+ \frac{1}{4}ax + \left( \frac{1}{4}b - \frac{3}{32}a^2 \right)x^2 +\left( -\frac{3}{16}ab+ \frac{7}{128}a^3 \right)x^3$.
}
,\\ $\cdot$ it well-behaves for $\Lambda\in \MB{R}$
, \\
$\cdot$ the contribution is same as the background subtraction method \cite{GibbonsHawking} for asymptotically flat spacetimes with an $S^{d-2}\times \MB{R}$ boundary topology. 

The explicit form of the corresponding Brown-York tensor is
\bea
\tau_{ij}^{proper} = \hspace{5.5cm} \notag \\
 \frac{-1}{8\pi G} \left[ \Theta_{ij}-\gamma_{ij}(\Theta-\Theta_{Mann}) -2 \frac{\pd \Theta_{Mann}}{\pd \gamma^{ij}}  \right] 
\ena
\bea
2 \frac{\pd \Theta_{Mann}}{\pd \gamma^{ij}}= \hspace{6.cm} \notag \\ \begin{cases}
    \displaystyle \hspace{3.cm} 0 \hspace{3.cm}  d=3 \\
    ~\\
    \displaystyle \hspace{1.8cm} \frac{1}{\Theta_{Mann}} C_{1}(d)\MC{R}_{ij} \hspace{1.8cm}  d=4,5 \\
    ~~ ~~ \\
    \displaystyle \frac{1}{2\Theta_{Mann}^3} \left[D_{1}(d)\left(2\MC{R}^{~k}_{i}\MC{R}_{kj} + \frac{d-7}{d-2} \MC{R}\MC{R}_{ij}  \right) \right.  ~ \\
   \hspace{4cm}  - D_{2}(d) \Lambda \MC{R}_{ij}  \biggl]  ~  d=6,7  
\end{cases} \notag
\ena

\section*{$S^{d-2} \times \MB{R}$ Boundary Geometries}
I show the utility of Mann counterterm for a class of spacetimes with an $S^{d-2} \times \MB{R}$ boundary geometry, whose metric form is
\bea
\BS{\gamma}= -d\ti{t}^2 + r_{b}^2 d\Omega_{d-2} ~ , \label{Boundary}
\ena
$r_{b}$ is the radius of $S^{d-2}$. The ground state of spacetimes with the boundary would be pure flat, dS, and AdS spacetime with a cut-off at $r=r_{b}$;
\footnotemark[8]\footnotetext[8]{
What I mean by ``ground state'' here is, in a classical mechanical sense, the most lowest energy solution with the boundary condition (\ref{Boundary}) and, in a quantum mechanical sense, the coarse-grained and decohered history of the ground state with a overwhelming probability (if exists) at some level of coarse-graining.  
}
\bea
\BS{g}^{gs}= -f(r)dt^2 + \frac{1}{f(r)}dr^2 + r^2 d\Omega_{d-2} \label{Bulk} \\
 f(r)=\left(1- \frac{2}{(d-2)(d-1)}\Lambda r^2  \right) \\
  r\in[0,r_{b}] \notag \hspace{1cm}
\ena
Only for the $\Lambda>0$ case, I restrict the analysis to $r_{b}< \sqrt{\frac{(d-2)(d-1)}{2\Lambda}}$.
The extrinsic curvature is
\bea
\Theta_{tt}= \frac{2\Lambda r_{b}}{(d-2)(d-1)}\sqrt{f(r_{b})} ~ , \hspace{2.6cm} \\
\Theta_{ti}=0 ~ , \hspace{5.8cm} \\
\Theta_{ab}= \frac{\sqrt{f(r_{b})}}{r_{b}}\sigma_{ab} ~, \hspace{4.2cm} \\
\Theta= \frac{1}{\sqrt{f(r_{b})}}\left[ -\frac{2\Lambda r_{b}}{(d-2)(d-1)} + \frac{(d-2)}{r_{b}}f(r_{b}) \right]
\ena
where $\sigma_{ij}$ is the projection of $\BS{\gamma}$ to $S^{d-2}$ and $a,b, \cdots$ is the index of a coordinate on $S^{d-2}$, that is, $\sigma_{it}=\sigma^{it}=0$. 
Since Ricci tensor $\MC{R}_{ij}$ and Ricci scalar $\MC{R}$ of the boundary metric (\ref{Boundary}) are
\bea
\MC{R}_{ij}=\frac{(d-3)\sigma_{ij}}{r_{b}^2}, ~~ \MC{R}= \frac{(d-3)(d-2)}{r_{b}^2} ~ ,
\ena
$\Theta_{Mann}$ and $\displaystyle 2 \frac{\pd \Theta_{Mann}}{\pd \gamma^{ij}}$ are calculated as
\bea
\Theta_{Mann}= \frac{(d-2)}{r_{b}}\sqrt{f(r_{b})} ~ , \label{CALCULATION1}  \\
2 \frac{\pd \Theta_{Mann}}{\pd \gamma^{ij}}= \frac{1}{r_{b}\sqrt{f(r_{b})}} \sigma_{ij} ~ , \label{CALCULATION2} 
\ena
for $4\leq d \leq 7$.
From these quantities, we can easily confirm that
\bea
\tau^{proper}_{ij}=0 ~ , \label{CALCULATION3} 
\ena
for $4\leq d \leq 7$, for any $r_{b}$ and for any $\Lambda$
\footnotemark[9]\footnotetext[9]{
Except $r_{b}> \sqrt{\frac{(d-2)(d-1)}{2\Lambda}}$ for $\Lambda>0$.
}
.
Another way to see this is to compute the free energy at zero temperature through the canonical partition function defined by Euclidean path integral \cite{GibbonsHawking}. The Euclidean saddle corresponding to thermal equilibrium at low temperature $1/\beta$ is obtained by Wick rotating $t=-i\tau$ the metric (\ref{Bulk}). The on-shell value of each term of the Euclidean version of  (\ref{theProperAction}) is 
\bea
I^{E}_{EH}[\BS{g}^{gs, E}]= \frac{-\beta \Omega_{d-2}}{4\pi G} \frac{\Lambda}{(d-2)(d-1)} \frac{r_{b}^{d-1}}{\sqrt{f(r_{b})}} ~ , \hspace{0.5cm} \\
I^{E}_{YGH}[\BS{g}^{gs, E}]=\frac{-\beta \Omega_{d-2}}{8\pi G}\left[\frac{d-2}{r_{b}^2} - \frac{2\Lambda}{d-2} \right]\frac{r_{b}^{d-1}}{\sqrt{f(r_{b})}} ~ , \\
I^{E}_{Mann}[\BS{\gamma}^{E}]= \frac{\beta \Omega_{d-2}}{8\pi G} (d-2)\sqrt{f(r_{b})}r^{d-3} ~ , \hspace{1.3cm} \label{CALCULATION4} 
\ena
where $\Omega_{n}$ is the volume of unit $n$ dimensional sphere. Then $F=\frac{1}{\beta} I^{E}_{GR, proper}[\BS{g}^{gs, E}]=0$
\footnotemark[10]\footnotetext[10]{
For the $d=4,5$ case, see also the calculation by \cite{KrausLarsenSiebelink}.
}
.
This subtraction is always valid for any spacetimes with an $S^{d-2}\times \MB{R}$ boundary geometry both of infinite and finite spatial volume
\footnotemark[11]\footnotetext[11]{
For asymptotically AdS spacetimes of odd dimension, there is known to exist the non-vanishing ground state energy if we compute it by using $I_{ct}$, that is interpreted as the Casimir energy of the Holographic CFT \cite{BalasubramanianKraus}. The existence is related to the ambiguity of the Holographic counterterm of odd dimension, that is, the freedom of adding curvature invariants of an appropriate mass dimension to $I_{ct}$. For large $l$ or small curvature limit, $I_{Mann}$ and $I_{ct}$ are differ by such invariants and the difference exactly compensates the Casimir energy for asymptotically AdS spacetimes with an $S^{d-2} \times \MB{R}$ boundary geometry.
}
.
Therefore, when we analyze standard gravitational thermodynamics \cite{GibbonsHawking, HawkingPage, York2, BrownComerMartinezMelmedWhitingYork}, all we need is to just add Mann counterterm $(\ref{MannTerm1})$.

\section*{Some Remarks}
\noindent
\underline{Higher dimension} \\

The extension to higher dimension is straightforward. For example, using Maclaurin expansion formula for $\left(1+x \right)^{\frac{1}{6}}$
\footnotemark[12]\footnotetext[12]{Explicitly, 
$
\left( 1+ax+bx^2 + cx^3 \right)^{\frac{1}{6}} \simeq 1 + \frac{1}{6}ax $
$+ \left(\frac{1}{6}b - \frac{5}{72}a^2 \right)x^2 $
\\
$
+ \left( \frac{1}{6}c - \frac{5}{36}ab + \frac{55}{1296}a^3 \right)x^3
$ \\
$
+ \left( -\frac{5}{36}ac - \frac{5}{72}b^2 + \frac{55}{432}a^2 b - \frac{935}{31104}a^4 \right)x^4 .
$
}
,  we can construct Mann counterterm for $d=8,9$ that coincides with the Holographic counterterm for $d=8,9$ presented in \cite{KrausLarsenSiebelink} for asymptotically AdS spacetimes (up to a finite contribution for $d=9$). We can check that eq. (\ref{CALCULATION1}), (\ref{CALCULATION2}), (\ref{CALCULATION3}), (\ref{CALCULATION4}), and $F=0$ is also hold for $d=8,9$. For more higher dimension, Mann counterterm can be constructed by using the formula for $(1+x)^{\frac{1}{2\left[\frac{d-2}{2} \right]}}$, would properly subtract the ground state energy for spacetimes with an $S^{d-2} \times \MB{R}$ boundary geometry, and can be used for gravitational thermodynamics of higher dimension.  \\
\noindent \\
\underline{$\Lambda$ as the parameter of GR} \\

Someone might feel strange about assigning zero energy to dS/AdS spacetimes enclosed by an $S^{d-2}\times \MB{R}$ boundary  (\ref{Bulk}) since the bulk can be thought as filled with a positive/negative energy if they regard $\Lambda$ as matter and consider the corresponding energy-momentum tensor $T_{\mu\nu}^{\Lambda}=\frac{-\Lambda}{8\pi G}g_{\mu\nu}$. Actually, from this point of view, their intuition would be correct. On the other hand, from another point of view, one can regard $\Lambda$ as just the parameter of the gravitational theory, and think (\ref{Bulk}) as ``empty'' dS/AdS spacetimes.   In this letter, I took the latter point of view and assigned zero energy to them by subtracting the ``fictitious energy'' of $\Lambda$, like we do for asymptotically AdS spacetimes in the context of AdS/CFT. Eventually, this matter is nothing but just difference of viewpoint, or choice of a reference of energy.

\noindent \\ 
\underline{Other boundary geometry and topology} \\

 Although Mann counterterm are shown to work for some class of spacetimes, it is far from saying that Einstein-Hilbert-York-Gibbons-Hawking-Mann action (\ref{theProperAction}) is the proper gravitational action. Pursuing the possibility is left for future work \cite{Miyashita}. Whether it is or not, it would be interesting to check whether Mann counterterm for $d\geq 6$ can eliminate not only the divergence of asymptotically flat spacetimes with an $S^{d-2}\times \MB{R}$ boundary topology but also that with another boundary geometry, such as $S^{n}\times \MB{R}^{d-1-n}$ (for $d\geq 6$), and see the relationship with the previously proposed counterterms for such spacetimes \cite{KrausLarsenSiebelink, MannMarolf}.

\noindent \\
\underline{Holography} \\

From the perspective of Holographic principle \cite{tHooft, Susskind}, adding Mann counterterm to the action of quantum gravity with a time-like boundary $(\MC{B}, \BS{\gamma})$ can be seen as adding it to the action of the Holographic QFT on $(\MC{B}, \BS{\gamma})$. Although the Holographic dual of gravity with a boundary of finite spatial volume is less-known compared to that of asymptotically AdS spacetimes, one promising thing may be that it is non-local \cite{LiTakayanagi}.  Suppose Mann counterterm leads to the proper gravitational action as well as the proper Holographic non-local QFT action. The different point from usual local field theories, such as the Holographic CFT, is that, for subtracting the ground state energy, Mann counterterm Lagrangian that is non-polynomial of curvature invariants is needed for the Holographic non-local QFT whereas a counterterm Lagrangian for local field theories, such as Mann counterterm Lagrangian for the Holographic CFT, that is polynomial of them is sufficient. At the moment, the meaning of this difference is unclear to me. I hope it could shed some light to general Holography and the Holographic non-local QFT.
~\\
~\\
\section*{Acknowledgement}
S.M. is grateful to Shintaro Sato for useful discussion. He also thanks Rodrigo Olea for helpful comments. This work was supported in part by a Waseda University Grant for Special Research Project (No. 2020C-775).

\end{document}